\documentclass[conference]{IEEEconf}

\IEEEoverridecommandlockouts                              % This command is only needed if 
\usepackage[english]{babel}
\usepackage{amsmath,amsfonts,amssymb, amsthm}
\usepackage{graphicx}
\usepackage[colorlinks=true, allcolors=blue]{hyperref}
\usepackage{parskip}
\usepackage{tabularx}
\usepackage{bm}
\usepackage{hhline}
\usepackage{multirow}
\usepackage{etoolbox}
\usepackage{cite}

\patchcmd{\thebibliography} %% removes spacing between bibliography items
  {\settowidth}
  {\setlength{\parsep}{0pt}\setlength{\itemsep}{0pt plus 0.1pt}\settowidth}
  {}{}

\newtheorem*{remark}{Remark}

\renewcommand{\t}{^{\mbox{\tiny\sf T}}} %% the transpose operator
\newcommand{\blkdiag}{\text{blkdiag}}

\renewcommand{\P}{\mathbb{P}}
\newcommand{\E}{\mathbb{E}}
\newcommand{\tr}{\text{tr}}

\def \b#1{\mathbf{#1}}

\allowdisplaybreaks

\title{\LARGE \bf Stochastic Control of UAVs: An Optimal Tradeoff between Performance, Flight Smoothness and Control Effort}

\author{George Rapakoulias$^{1}$ and Panagiotis Tsiotras$^{2}$
\thanks{$^{1}$ Ph.D. student, School of Aerospace Engineering, Georgia Institute of Technology, Atlanta, GA, 30332, USA. Email:
 {\tt\small grap@gatech.edu}}%
\thanks{$^{2}$ David and Andrew Lewis Chair and Professor, School of Aerospace Engineering, and Institute for Robotics and Intelligent Machines, Georgia Institute of Technology, Atlanta, GA, 30332, USA. Email:
 {\tt\small tsiotras@gatech.edu}}%
}

\begin{document}

\maketitle

\begin{abstract}
Safe and accurate control of unmanned aerial vehicles in the presence of winds is a challenging control problem due to the hard-to-model and highly stochastic nature of the disturbance forces acting upon the vehicle.
To meet performance constraints, state-of-the-art control methods such as Incremental Nonlinear Dynamic Inversion (INDI) or other adaptive control techniques require high control gains to mitigate the effects of uncertainty entering the system.
While achieving good tracking performance, IDNI requires excessive control effort, results in high actuator strain, and reduced flight smoothness due to constant and aggressive corrective actions commanded by the controller. 
In this paper, we propose a novel control architecture that allows the user to systematically address the trade-off between high authority control and performance constraint satisfaction.
Our approach consists of two parts. 
To cancel out biases introduced by unmodelled aerodynamic effects we propose a hybrid, model-based disturbance force estimator augmented with a neural network, that can adapt to external wind conditions using a Kalman Filter. 
We then utilize state-of-the-art results from Covariance Steering theory, which offers a principled way of controlling the uncertainty of the tracking error dynamics.
We first analyze the properties of the combined system and then provide extensive experimental results to verify the advantages of the proposed approach over existing methods.  
\end{abstract}

\section{Introduction}
During the last decade, the surge in research in small unmanned aerial vehicles (UAVs), especially quadrotors, has demonstrated their excellent capabilities for performing complicated tasks in harsh environments.
To perform these tasks successfully, a variety of different controllers have been developed and studied extensively.
For trajectory tracking tasks, state-of-the-art systems usually use a backstepping architecture where multiple nested control loops are used to control plant dynamics evolving at different time scales while exploiting differential flatness \cite{tal2020accurate}. 
High-performance systems also feature some method for compensating unmodelled dynamics, such as aerodynamic effects, and rely either on neural networks \cite{bauersfeld2021neurobem, o2022neural, cao2017gaussian} or adaptive control methods \cite{hanover2021performance, tal2020accurate}.

Regarding the different techniques used in the literature for compensating for aerodynamic disturbances, one design parameter is whether these methods assume some structure for aerodynamic effects, either in the form of an analytic model or a neural network trained on flight data. 
Another design parameter is whether these algorithms provide adaptation capabilities with regard to the ambient wind conditions. 
Canceling out aerodynamic effects without an aerodynamic model leverages the fact that these usually evolve at a slower timescale, and are directly observable. These methods include \cite{tal2020accurate, hanover2021performance, smeur2016adaptive, paris2020dynamic} and since they make no modeling assumptions, can work with ambient wind, however, they require high-end, usually fully customized hardware and sensors running at very high sampling rates to achieve high performance.
On the other hand, works that achieve high-performance trajectory tracking but cannot adapt with respect to external wind include \cite{bauersfeld2021neurobem, cao2017gaussian, torrente2021data, kaufmann2023champion} and leverage machine learning. 
Model-based approaches with adaptive capabilities include \cite{xing2023active}, which uses a computationally heavy particle filter for the estimation of wind conditions, and \cite{davoudi2023physics}, where an ultrasonic wind sensor is required to estimate relative windspeed. Another recent work aiming to bridge the gap between adaptation capabilities and machine learning is \cite{o2022neural}, where the authors developed a neural network-based adaptive controller that only requires the online estimation of a small number of basis function weights to characterize external wind conditions. 

Another way to differentiate existing control architectures is with regard to how they handle uncertainty. Most algorithms use linear feedback with static gains for the outer position control loop and are usually tuned using Linear Quadratic Regulator theory or through extensive experimentation. 
While these methods work well, they are usually tuned in a task-dependent manner. 
High authority control, is used in cases where accurate trajectory tracking is required, at the expense of aggressive attitude corrections, while lower values of control gains yield a smoother flight at the expense of tracking accuracy.

In this paper, a new control architecture based on Optimal Covariance Steering (OCS) theory \cite{liu2022optimal, rapakoulias2023discrete, okamoto2018optimal, bakolas2018finite} is investigated and tested for controlling quadrotors in the presence of stochastic disturbances. 
Our approach combines two methods.
The first method focuses on converting the dynamics of the quadrotor to those of a 3D double integrator subject to stochastic disturbances. 
This is done by using standard algorithms for quadrotor attitude control, along with a new aerodynamic force estimator based on a Kalman Filter and a composite analytic/neural network-based aerodynamic drag model.
The technique is similar to Nonlinear Incremental Dynamic Inversion (INDI) but replaces the low-pass filters of acceleration measurements with a model-based optimal estimator, yielding less noisy disturbance estimates. 

The second step utilizes results from the Optimal Covariance Steering theory to efficiently control the uncertainty of the system. 
Contrary to traditional stochastic control approaches, which aim at controlling the trajectory of a stochastic system in the presence of uncertainty, OCS theory aims at driving the distribution of the state to an explicitly defined target distribution, while respecting probabilistic constraints.
This research field stems from a wider and rapidly evolving area of control theory, aiming at controlling the state distribution of dynamical systems rather than the evolution of a mean trajectory, and finds applications to mean field and multiagent control \cite{ruthotto2020machine, chen2023density, Saravanos-RSS-21, saravanos2023distributed, rapakoulias2024discrete}, stochastic maximum coverage control \cite{aggarwal2024sdp}, and safe stochastic model predictive control \cite{knaup2023safe, saravanos2022distributed}, among many others.
% Covariance Steering is a stochastic optimal control technique that aims to control the state distribution of the system by imposing various performance specifications as probabilistic constraints on the error dynamics. 
In practical applications, covariance control techniques have been shown to improve control performance in systems with stochastic dynamics such as high-performance off-road autonomous driving \cite{knaup2023safe, yin2022trajectory} and Planetary Entry, Descent, and Landing applications \cite{ridderhof2018uncertainty, selim2024Stochastic}. 
In the context of VTOL UAV control, we argue that OCS offers a principled way of trading off high-gain feedback control to achieve a smoother and more fuel-efficient flight.

After developing the theoretical framework of the proposed approach, we benchmark our method by conducting comprehensive experiments in two tasks: trajectory tracking of aggressive maneuvers and landing, both in the presence of strong turbulent winds.
Compared to the state-of-the-art, our method results in smoother flight and smaller control signals, while respecting performance constraints.

\section{Quadrotor equations of motion}
The problem of modeling the dynamics of a quadrotor has received a lot of attention in the literature and therefore it will not be discussed in detail here. We refer the reader to \cite{mellinger2011minimum} and the references therein. A brief summary is included below for the sake of completeness. 
Although different levels of detail regarding the dynamics can be considered, the most common model for position control is that of a point mass.

To this end, consider a point mass $m$ attached at the origin of a frame $\mathcal{B}$. 
The rotation matrix of $\mathcal{B}$ w.r.t. the inertial frame  $\mathcal{I}$ is denoted by $R$ and is parametrized with the unit quaternion $\mathbf{q}$. 
The equations of motion are 
\begin{subequations}
    \begin{align}
        & m \ddot{\b r} = m \b g  - R \hat{\b e}_z \tau + \b f_d \label{plant1}, \\
        & \dot R = R \omega^{\times},
    \end{align}
\end{subequations}
where $\hat{\b e}_3$ is the unit vector in the vertical z inertial direction, $\tau$ is the total thrust and $\b f_d$ is a disturbance force that is mostly due to unmodelled aerodynamic effects such as drag. Throughout this paper, it is assumed that the quadrotor has a low-level control loop running at high frequency allowing it to follow orientation commands, i.e., any reference attitude $\b q_r$ and any thrust setpoint $\tau_r$. 
For the hardware used in our experiments (the PX-4 autopilot \cite{PX4}) the default attitude control loop implemented on PX-4 was found to be sufficient for the task of reference tracking and was used without any modifications.
This allows us to write the equations of motion in the form 
\begin{equation}\label{double_int}
    m \ddot{\b r} = m \b g  + \b f_c  + \b f_d,
\end{equation}
where $\b f_c$ is the new, high-level control input that corresponds to a 3D control force in any direction, not just in the thrust axis. 
The attitude and thrust setpoints can be calculated by up to a heading constant \cite{mellinger2011minimum}, while the latter one can be selected arbitrarily depending on the application. Specifically, the relevant equations are
\begin{subequations}
    \begin{align*}
        & \tau_d = \| \mathbf{f_c} \|, \quad \b x_c = [1 \quad 0 \quad 0]\t, \\
        & \b z_b = \mathbf{f_c} / \| \mathbf{f_c} \|, \quad \b y_b = \frac{\b z_b \times \b x_c}{ \| \b z_b \times \b x_c \|}, \quad \b x_b = \b y_b \times \b z_b ,  \\
        & R_d = [\b x_b \quad \b y_b \quad \b z_b].
    \end{align*}
\end{subequations}
We set the control force $\b f_c$ equal to 
\begin{equation} \label{control_law}
    \b f_c = -m \b g - \hat{\b f}_d + m \b u
\end{equation}
where $\hat{\b f}_d$ is the estimated aerodynamic force. Assuming that the latter one is accurate, with an error that corresponds to a zero mean Gaussian random variable i.e. $\hat{\b f}_d = \b f_d + \b \epsilon$ where $\epsilon$ is a zero mean Gaussian process with known covariance $Q_w$. 

\section{Drag compensation algorithm}
When the quadrotor flies at high speeds, various aerodynamic forces act upon it. 
The biggest influence is the body and rotor drag forces, as well as the change of the motor's thrust curve due to the effect of the incoming wind \cite{bauersfeld2021neurobem}. 
We will focus only on the first phenomenon since it has the largest effect on flight performance as far as trajectory tracking, which is also found to be the case in \cite{o2022neural}.

In order to use a model-based method for drag compensation, a brief overview of existing drag models is presented first.
The simplest studied analytic drag models for quadrotors is the linear model \cite{sikkel2016novel}, while more complicated methods include the effects of a possibly asymmetric body \cite{faessler2017differential} and heuristic modifications for including rotor drag \cite{svacha2017improving} and blade flapping \cite{huang2009aerodynamics}. 
More complicated analytic models have also been studied, but usually in the context of simulation rather than for control. 
For example in \cite{pascasio2020rapid} a physics-based model is proposed using interpolated drag coefficients calculated from an accurate CFD model to capture static drag effects and a dynamic model based on an unsteady aerodynamic analysis to captures transient effects.
As an alternative to analytic models, recently proposed methods based on neural networks have shown great effectiveness in capturing the physics of the problem \cite{o2022neural, bauersfeld2021neurobem}.
Although these methods boast high performance in reference conditions, they purely rely on machine learning techniques rather than on fundamental aerodynamic principles to capture aerodynamic effects and therefore require a large amount of training data, usually gathered from wind tunnel testing, to adequately sample the behavior of the quadrotor in a sufficiently wide range of flight conditions.
Furthermore, because drag estimates are used in closed-loop, analyzing the stability of such systems poses further difficulties. 
Seminal works addressing the issue include \cite{shi2019neural, o2022neural}. 

Our method aims to provide a solution leveraging the advantages of a model-based technique and the flexibility of machine learning approaches.
Specifically, we employ a Kalman Filter-based adaptive technique similar to \cite{o2022neural} but instead of using a purely neural network-based model, we combine it with a linear drag model. 
Experimental data validate that the linear drag model captures most of the physics of the problem while the inclusion of the neural network part increases performance by capturing higher-order effects.

The model we consider is linear with respect to the relative wind of the quadrotor, but nonlinear with respect to the attitude and rotor speed. Specifically, we assume
\begin{equation}\label{drag_model}
    \hat{\b f}_d = \underbrace{\b C_d (\b w - \b v)}_{\text{linear model}} + \underbrace{\b \Phi (\b q, \Bar{\eta}) (\b w - \b v)}_{\text{NN model}},
\end{equation}
where $\b C_d = \blkdiag(C_{x}, C_{y}, C_{z})$ and $\b \Phi (\b q, \Bar{\eta})$ is a ReLU Neural Network (NN) taking the attitude $\b q$ and motor RPM $\Bar{\eta}$ as inputs. Its architecture consists of 4 layers and $\{8, 20, 20,3\}$ nodes per layer.
The model \eqref{drag_model} is symmetric with respect to the body frame absolute velocity $\b v$ and wind velocity $\b w$, making it compatible with the physics of the problem. 
This last remark allows us to train it with data only from zero ambient wind conditions where the drag forces are only due to the inertial velocity term $\b v$. 
The training of the model is performed on PyTorch \cite{paszke2017automatic} using the ADAM optimizer and Thikonov regularization to mitigate overfitting \cite{bishop2006pattern}. 
%
% \begin{figure}[!h]
%     \centering
%     \includegraphics[width=1\linewidth]{figures/correlation_Lin_NN.png}
%     \caption{Correlation between drag ground truth and linear model, Neural Network model}
%     \label{fig:Correlation plot}
% \end{figure}

To adapt the model online with respect to $\b w$, one can either use custom sensors to measure the windspeed directly, as in \cite{davoudi2023physics, xing2023active}, or use the fact that drag forces are directly observable through equation \eqref{plant1}, where each quantity can be directly measured or estimated, except $\b f_b$. 
Specifically, in our implementation, we estimated acceleration using IMU measurements averaged over the sampling time of the adaptation algorithm, while for thrust estimates we used RPM measurements in combination with the experimentally measured thrust curve of the motor. 
Therefore, one can adapt the drag model online to ensure it matches the noisy drag estimates from equation \eqref{plant1}.
For the adaptation law, we use an Extended Kalman Filter (EKF) due to its small computational requirements, making it suitable for online use in CPU-constrained hardware such as the offboard computer of a small drone. 
We also highlight that a key parameter for the effectiveness of the method is fast adaptation rather than a very accurate drag model, or complicated adaptation algorithms such as particle filters. Specifically, the EKF update equations are:
\begin{subequations}
    \begin{align}
        % & \hat{\b f}_{d, k|k-1} = \b \Phi(\b w_k , \b x_k) \\
        & \b P_{k|k-1} = \b P_{k-1|k-1} + \b Q, \\
        & \b K_k = \b P_{k|k-1} \b H_k \t (\b H_k \b P_{k|k-1} \b H_k \t + \b R_k)^{-1}, \\
        & \hat{\b f}_{d, k|k} =  \hat{\b f}_{d, k|k-1} + \b K_k ( \b f_d - \hat{\b f}_{d, k|k-1}),
    \end{align}
\end{subequations}
% \begin{subequations}
%     \begin{align}
%         & \dot{\hat{\b w}} =  \b K ( \b f_d - \hat{\b f}_d) \\
%         & \b K_k = \b P \b H \t \b R^{-1} \\
%         &  \dot{ \b P} = - \b P \b H\t \b R^{-1} \b H \b P + \b Q
%     \end{align}
% \end{subequations}
%
where $\hat{\b f}_d$ is the model-predicted drag and 
\begin{equation}
    \b H_k = \frac{\partial \hat{\b f}_d }{\partial \b w} = \b C_d + \b \Phi(\b q_k, \Bar{\eta}_k)
\end{equation}
is the Jacobian of the drag model with respect to the external wind, and the index $k$ corresponds to the discrete measurements at the current time instant. The convergence of the filter owns to the linear parametrization with respect to the estimated parameters and the stability of the standard linear Kalman Filter.

\begin{figure}
    \centering
    \includegraphics[width=1\linewidth]{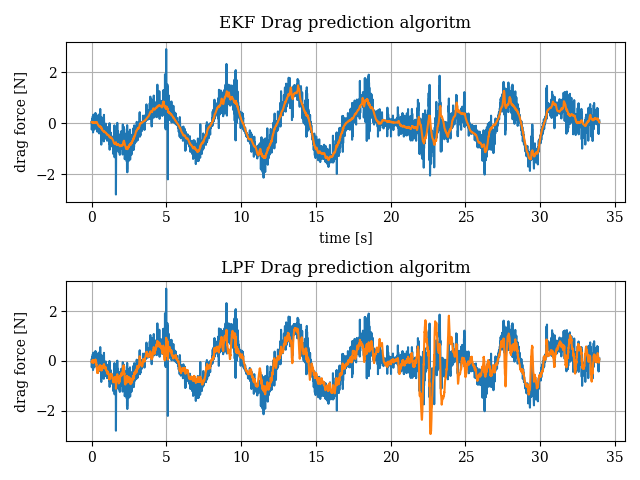}
    \caption{Model-based EKF filtering vs INDI first order LP filter with $f_c = 5$ Hz. Blue line corresponds to raw data and orange to filtered data.}
    \label{fig:drag_prediction}
\end{figure}

Figure \ref{fig:drag_prediction} shows the predicted drag forces using the proposed approach compared to the low pass filter approach proposed in INDI \cite{tal2020accurate}. 

\begin{remark}
    Compared to INDI, this method replaces the low pass filters used to attenuate acceleration and RPM measurement noise of the drag forces from equations \eqref{plant1} with a model-based optimal estimator. 
    This results in less noisy drag estimates without adding the lag of a very low-frequency low-pass filter or requiring very high sampling rate sensors. 
    Compared to the method in \cite{o2022neural}, we use only one basis function for the drag estimation, requiring the online estimation of only three parameters, that is, the components of $\b w$, instead of twelve, and therefore resulting in faster adaptation. 
    Finally, the symmetry of the model with respect to the relative wind speed allows us to train it without requiring hard-to-acquire wind tunnel data. 
\end{remark}

\begin{figure}
    \centering
    \includegraphics[width=1\linewidth]{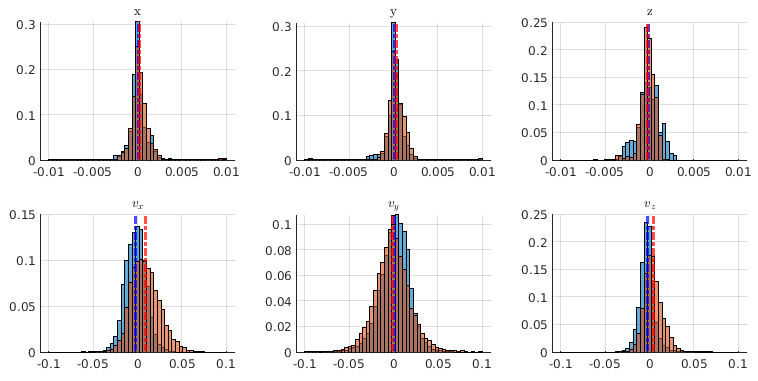}
    \caption{Disturbance distribution before(red) and after(blue) aerodynamic force compensation for a flight into 7 $m/s$ wind in the x-direction. The mean error is represented by dashed lines.}
    \label{fig:err_distr}
\end{figure}

The performance of the method is evaluated visually in Figures \ref{fig:drag_prediction} and \ref{fig:err_distr}. Figure \ref{fig:drag_prediction} shows cleaner drag estimates compared to the low pass filter of the standard INDI controller. Figure \ref{fig:err_distr} shows the modeling error distribution of the double integrator model coming from substituting \eqref{control_law} in \eqref{double_int}, evaluated from experimental data, for a flight into 7 m/s wind, with and without the aerodynamic compensation. Overall, the compensator reduces the uncertainty and eliminates biases.

\section{Optimal Covariance Steering control law}

Optimal Covariance Steering (OCS) is a finite-horizon stochastic optimal control method used to control the state distribution of a linear system.
The formulation used in this paper is
\begin{subequations} \label{reference_problem}
\begin{align}
& \min_{x_k, u_k} \quad J =\E [\sum_{k = 0}^{N-1} { x_k\t Q_k x_k + u_k \t R_k u_k}], \label{ref:cost}
\end{align}
such that, for all $k = 0, 1, \dots, N-1$,
\begin{align}
& x_{k+1} = A_k x_k + B_k u_k + D_k w_k, \label{ref:dyn} \\
% & \E[x_1] = \mu_i \label{ref:in_mean} \\
% & \cov(x_1) = \Sigma_i  \label{ref:in_cov} \\
% & \E[x_N] = \mu_f \label{ref:final_mean} \\
% & \cov(x_n) = \Sigma_f \label{ref:final_cov} \\
& x_0 \sim \mathcal{N}(\mu_i, \Sigma_i), \label{ref:initial_distr} \\
& x_N \sim \mathcal{N}(\mu_f, \Sigma_f), \label{ref:final_distr} \\
& \P(x_k \in \mathcal{X}) \geq 1-\epsilon_1, \label{ref:chan_con_x}\\
& \P(u_k \in \mathcal{U}) \geq 1-\epsilon_2 . \label{ref:chan_con_u}
\end{align}
\end{subequations}
In practice, usually, a state feedback control law 
\begin{equation} \label{linear_feedback}
    u_k = K_k(x_k - \mu_k) + v_k,
\end{equation}
where $\mu_k = \E[x_k]$ is assumed \cite{liu2022optimal, rapakoulias2023discrete}, converting problem \eqref{reference_problem} to a finite-dimensional optimization problem aiming to calculate the optimal gains $K_k$ and feedforward actions $v_k$.

Assuming the linear feedback law \eqref{linear_feedback}, Gaussian disturbances $w_k \sim \mathcal{N}(0, \Sigma_w)$ and affine polytopic expressions of the form 
\begin{subequations}
    \begin{align*}
        & \P ( \alpha_{x, i}\t x \leq b_{x, i} ) \geq \delta_x \\
        & \P ( \alpha_{u, i}\t x \leq b_{u, i} ) \geq \delta_u, \; \text{for} \quad i = 1, 2, \dots, n_c,
    \end{align*}
\end{subequations}
for \eqref{ref:chan_con_u}, \eqref{ref:chan_con_x}, Problem \eqref{reference_problem} can be reformulated as a linear semidefinite program as follows 
\begin{subequations} \label{convex_eq}
\begin{align}
& \min_{\Sigma_k, U_k, Y_k} \quad J_\Sigma = \sum_{k = 0}^{N-1} {\tr \big(Q_k \Sigma_k \big) + \tr \big(R_k Y_k \big)} \\
& \qquad \qquad \qquad \qquad  + \mu_k \t \bar{Q} \mu_k + v_k \t \bar{R} v_k, \nonumber
\end{align}
such that, for all $k = 0, 1, \dots, N-1$,
\begin{align}
& A_k \Sigma_k A_k\t + B_k U_k A_k\t + A_k U_k\t B_k\t + B_k Y_k B_k\t \nonumber \\
& \qquad + D_k D_k\t - \Sigma_{k+1} = 0,\label{convex_eq:cov_dyn} \\
& U_k \Sigma_k^{-1} U_k\t - Y_k \preceq 0, \label{convex_eq:relaxation} \\
& \Sigma_N  \preceq \Sigma_f, \label{convex_eq:final_cov} \\
& \mu_{k+1} = A_k \mu_k + B_k v_k, \\
& \ell \t \Sigma_k \ell + \alpha_x\t \mu_k - \beta_x \leq 0, \label{convex_constrained_cov} \\
& e\t Y_k e + \alpha_u \t v_k - \beta_u \leq 0.        \label{convex_constrained_effort} 
\end{align}
\end{subequations}
where,
\begin{subequations}
    \begin{align*}
        & \ell     = \sqrt{ \frac{\Phi(\delta_x)}{2 s_0} } \alpha_x, \quad s_0 = \sqrt{\alpha_x \t \Sigma_0 \alpha_x }, \\
        & \beta_x  = b_x - \Phi(\delta_x)  \frac{s_0}{2}  b_x, \\
        & e        = \sqrt{ \frac{\Phi(\delta_u)}{2 y_0} } \alpha_u, \quad y_0 = \sqrt{\alpha_u \t Y_0 \alpha_u }, \\
        & \beta_u  = b_x - \Phi(\delta_u)  \frac{y_0}{2}  b_u,
    \end{align*}
\end{subequations}

and $\Sigma_0, Y_0$ are linearization constants. For the detailed analysis, we refer the reader to \cite{rapakoulias2023discrete}. 
In all subsequent problems, formulation \eqref{convex_eq} is used.
Notice that in the cost function, we use different penalties for penalizing the mean state and control effort and their respective covariances.
This is because in a tracking problem, we are aiming to minimize the deviation from a nominal trajectory rather than penalize absolute values of the state. With this rationale, $Q_k$ is expected to have larger values than $\bar{Q}_k$, while the latter could even be omitted if penalizing absolute values of the state is not of interest. 

\begin{figure}[ht!]
    \centering
    \includegraphics[width=0.8\linewidth]{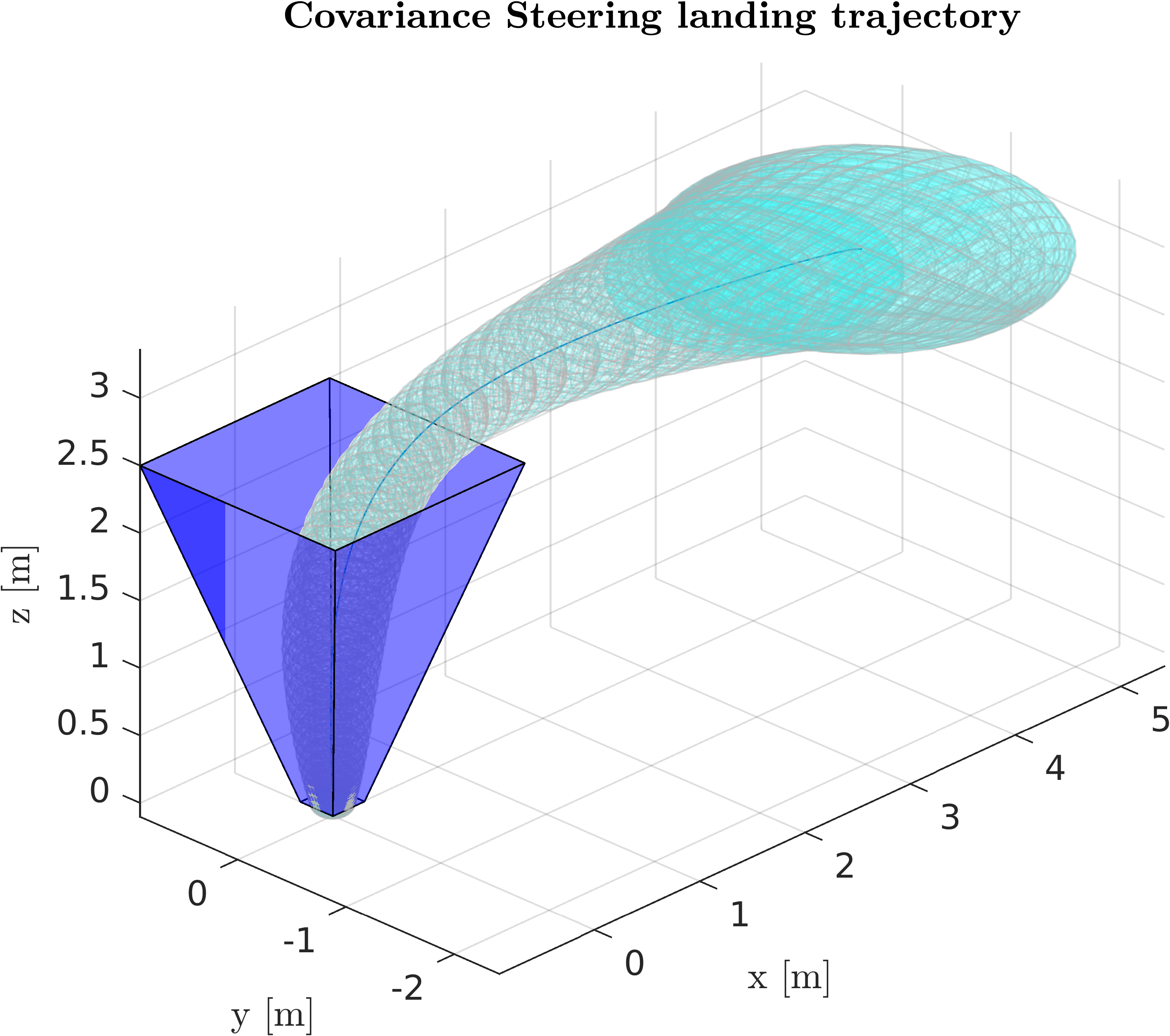}
    \caption{Optimal covariance steering landing problem solution with cone chance constraints to limit the feasible domain of the state. The blue ellipsoids represent the 3-$\sigma$ bounds for the state distribution and the blue line represents the mean trajectory. The dark blue conic surfaces represent chance constraints.}
    \label{fig:OCS_land}
\end{figure}

To illustrate our method, the solution of a covariance steering problem for a landing scenario is illustrated in Figure \ref{fig:OCS_land}. In order to constrain the approach of the vehicle to the landing zone, four chance constraints corresponding to each side of a cone are employed, written in the form of \eqref{convex_constrained_cov} with suitable parameters for $\ell, \; \alpha_x, \; \beta_x$. To control the landing position accuracy, an inequality terminal covariance constraint of the form \eqref{convex_eq:final_cov} is used.

\section{Experimental results}
\begin{figure}
    \centering
    \includegraphics[width=1\linewidth]{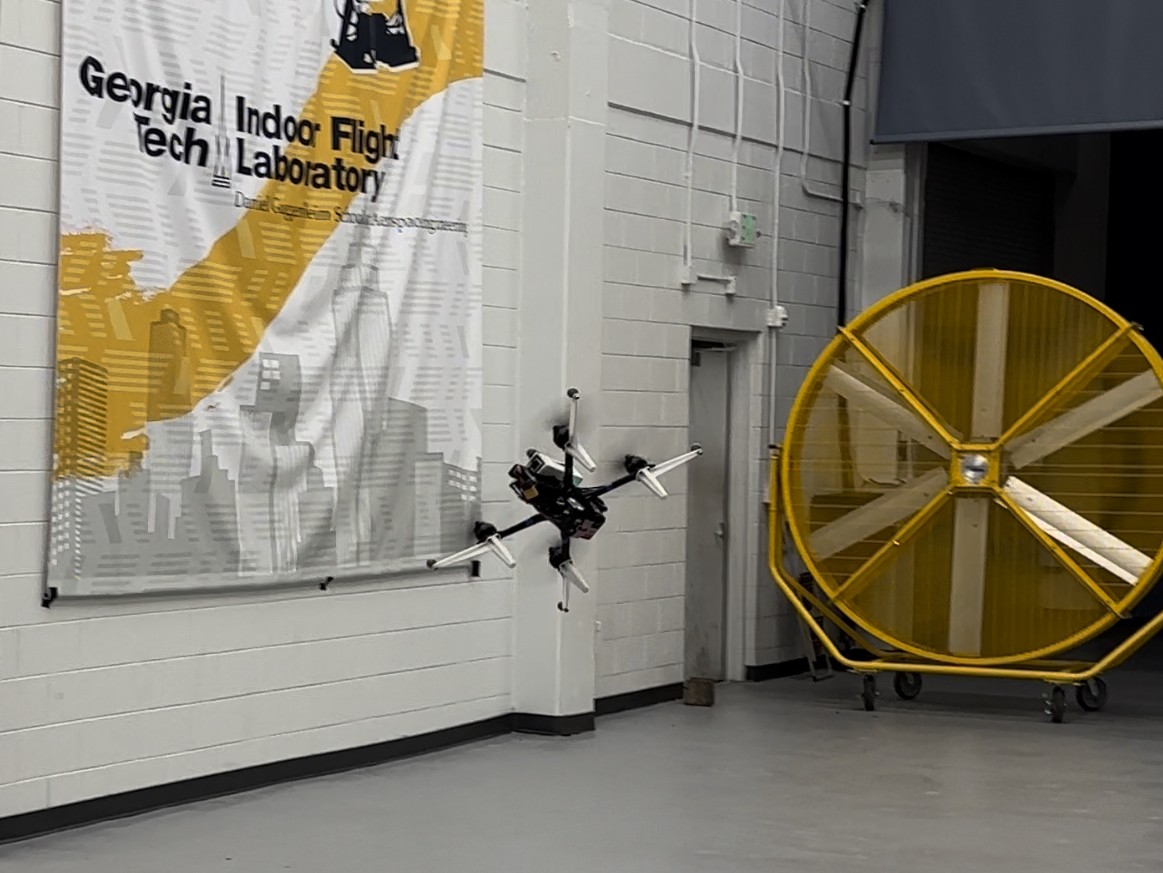}
    \caption{Quadrotor flying in the Georgia Tech's Indoor Flight Laboratory}
    \label{fig:setup}
\end{figure}

The equipment used for the experiments consists of a 5-inch racing quadrotor running PX-4 software with a total mass of 680 gr, and a maximum thrust of 39 N. 
The quadrotor is equipped with a RockPi-4B offboard computer running Ubuntu 20.04 with all programs being executed as ROS2 nodes. 
Position feedback is provided to the quadrotor from an external vision system at a rate of 100 Hz, and is fused with the onboard measurements on the flight controller's EKF.
Motor RPM feedback is obtained from the telemetry channel of the electronic speed controller, without using custom optical sensors. 
The setup is shown in action in Figure \ref{fig:setup}. For generating external wind, the testing facility features a large industrial fan capable of generating turbulent wind of speeds up to 3-4  m/s, as well as an array of leafblowers, capable of generating narrow airstreams of up to 10 m/s.

\subsection{Trajectory tracking}

\begin{figure}[ht!] 
    \centering
    \includegraphics[width=1\linewidth]{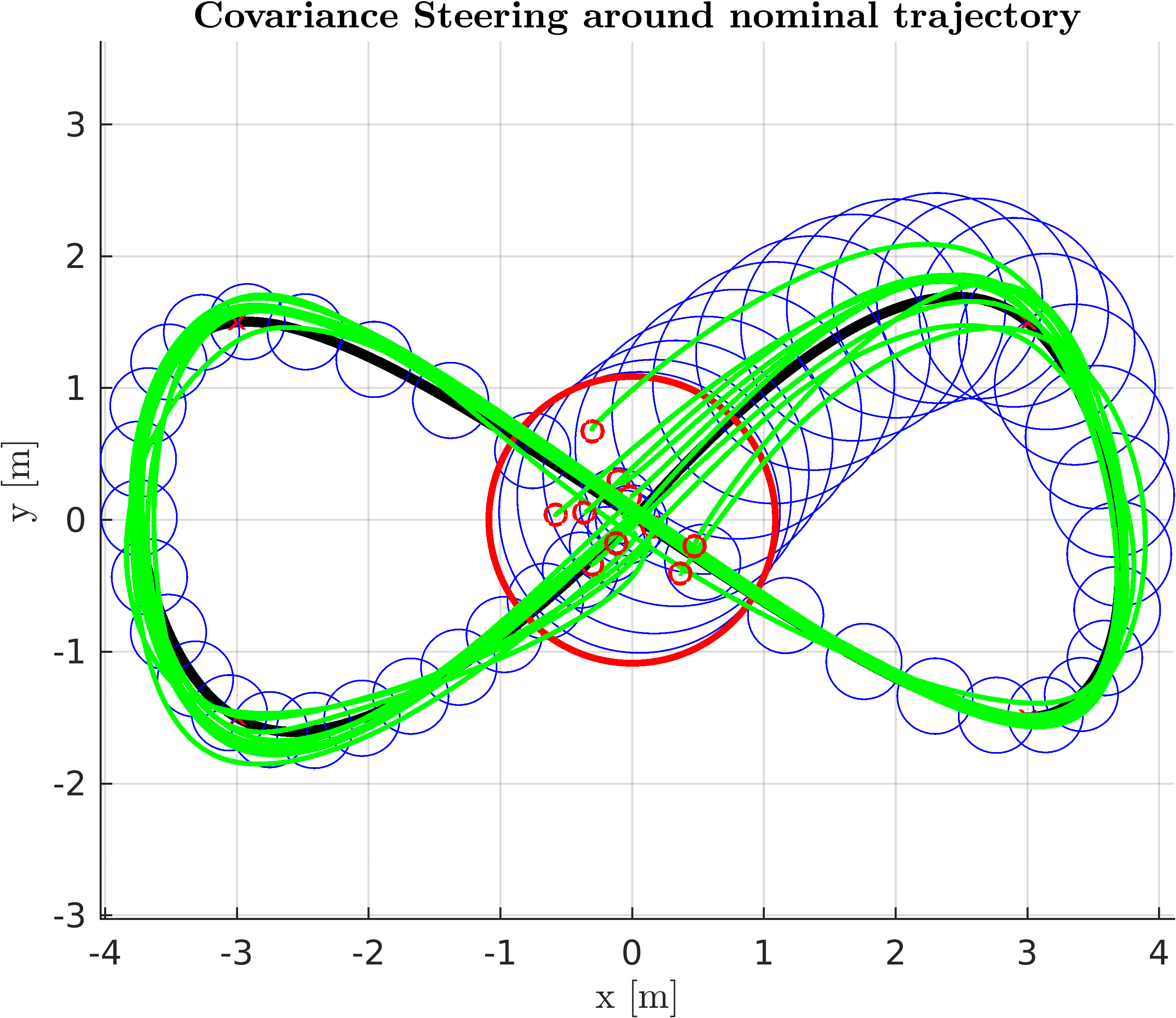}
    \caption{Figure-8 trajectory with experimental trajectories shown in green. Initial conditions are randomly sampled from the initial distribution of the covariance steering problem which is illustrated as the red circle centered in the origin. The $3-\sigma$ bounds are illustrated as blue ellipses.}
    \label{fig: fig-8-traj}
\end{figure}

\begin{figure}[ht!] 
    \centering
    \includegraphics[width=1\linewidth]{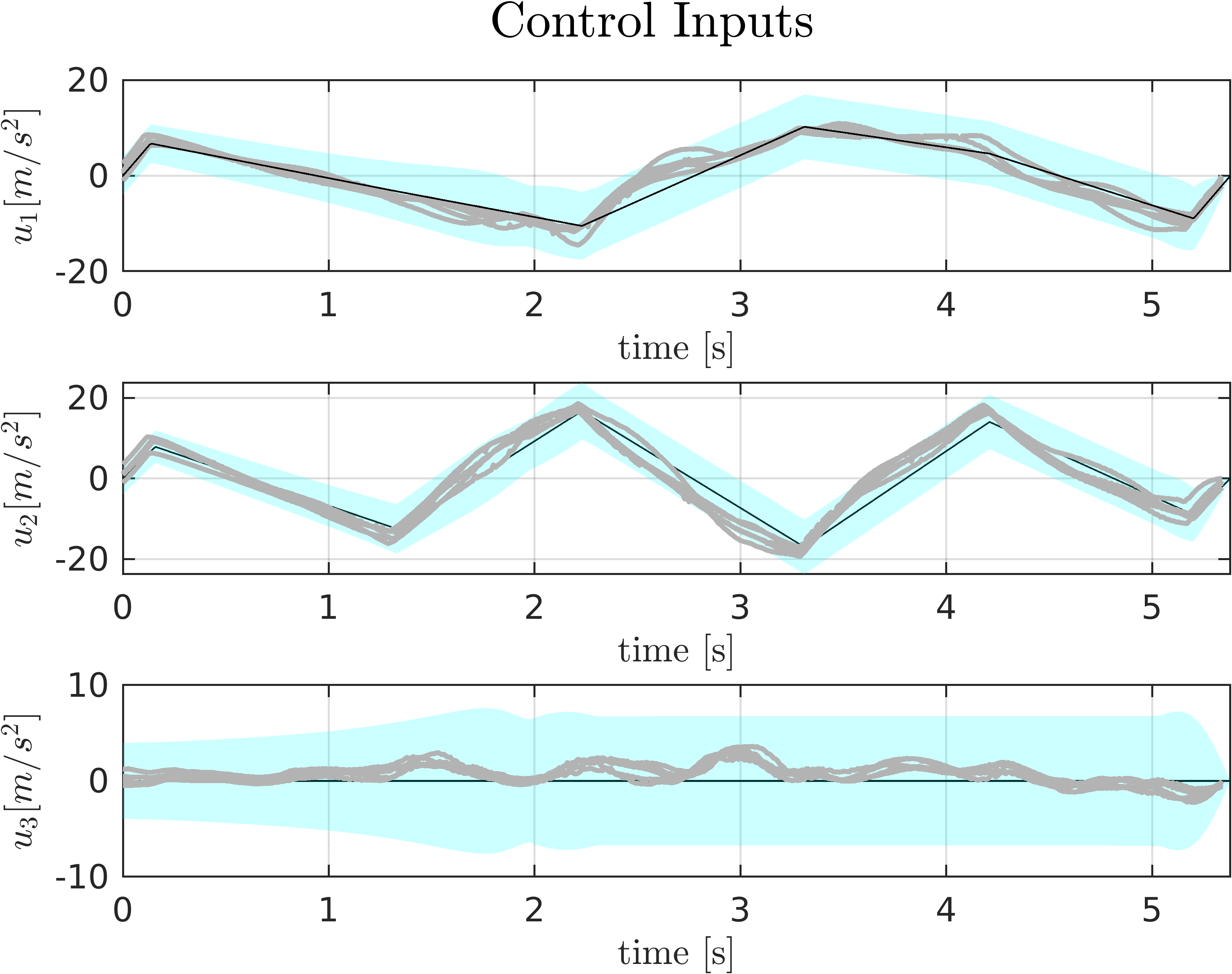}
    \caption{Figure-8 optimal control policy. The black lines correspond to the nominal, feedforward control action while the gray lines the actual, experimental control effort after the feedback term. The light blue area corresponds to the 3-$\sigma$ bounds calculated by the algorithm.}
    \label{fig:fig-8-inputs}
\end{figure}

First, the results regarding trajectory tracking of an aggressive figure-8 maneuver will be discussed. 
To evaluate the method, the trajectory shown in Figure \ref{fig: fig-8-traj} is considered. 
The control policy is calculated by solving a covariance steering problem of the form \eqref{convex_eq} with the addition of partial inequality covariance constraints on the first three states throughout the trajectory and four waypoint constraints on the mean trajectory. 
The optimal mean trajectory has a maximum speed of 12 m/s and and max acceleration of 22 m/s$^2$. 
Specifically, the problem parameters are
\begin{subequations}
\begin{align*}
    & A = \begin{bmatrix} \ I_2 & \Delta T I_2 \\ 0_2 & I_2 \end{bmatrix}, \quad B = \begin{bmatrix} 0_2 \\ \Delta T I_2  \end{bmatrix}, \quad \Delta T = 0.01, \\
    & D = \blkdiag(0.01 \; I_3, 0.1 \; I_3)
\end{align*}
\end{subequations}
and $Q = I_6, \; \bar{Q} = 0_3 \; \bar{R} = R = I_3$ .The total number of time steps is $N=540$. To limit the dispersion of the state and control effort from the mean values partial covariances constraints of the form
\begin{subequations}
\begin{align*}
& L\t \Sigma_k L \preceq (\delta_x/3)^2 \Sigma_c, \quad L\t = \begin{bmatrix} I_2 & 0_2 \end{bmatrix},\; \delta_x = 0.025 \;, \\
& Y_k \preceq (\delta_u/3)^2 I_3, \quad \delta_u = 10
\end{align*}
\end{subequations}
are added to the state and control effort covariances for time steps after $k\geq 100$.  The parameters $\delta_x$ and $\delta_u$ can be fine-tuned by the user parameters and correspond to the 3-sigma bounds for position and control effort respectively. 
The resulting covariance steering problem solution, overlayed with ten experimental trajectories is shown in Figure \ref{fig: fig-8-traj} while the corresponding control policy illustrated in Figure \ref{fig:fig-8-inputs}.
A quantitative comparison with the state-of-the-art can be viewed in Table \ref{tab:fig8_comparison}. 
Specifically, we use the RMS value of the tracking error as a metric for performance, and the RMS value of the angular acceleration as a metric for flight smoothness.
The benefit of the method stems from the fact that it provides a tool to systematically trade off tracking performance for a smoother flight and use of less control effort while also ensuring that performance constraints are satisfied.
\begin{table}[ht!] 
    \centering
    \caption{Comparison with state-of-the art for the trajectory tracking benchmark.}
    \begin{tabular}{|c|c|c|c|}
    \hline
        & RMS $\| \b x - \b x_{ref} \|_2$ cm & RMS $ \| \dot{\b \omega} \|_2$ rad/s$^2$ \rule{0pt}{2.4ex} \rule[-1.2ex]{0pt}{0pt} \\ \hline
         OCS + EKF        & 5.9     & 17.1              \\ \hline
         OCS + IDNI       & 5.3     & 18.6              \\ \hline
         LQR + EKF        & 4.9     & 19.0              \\ \hline 
         LRQ + INDI       & 7.3     & 23.2              \\ 
    \hline 
    \end{tabular}
    \label{tab:fig8_comparison}
\end{table}

\subsection{Landing}

\begin{figure}[ht!]
    \centering
    \includegraphics[width=1\linewidth]{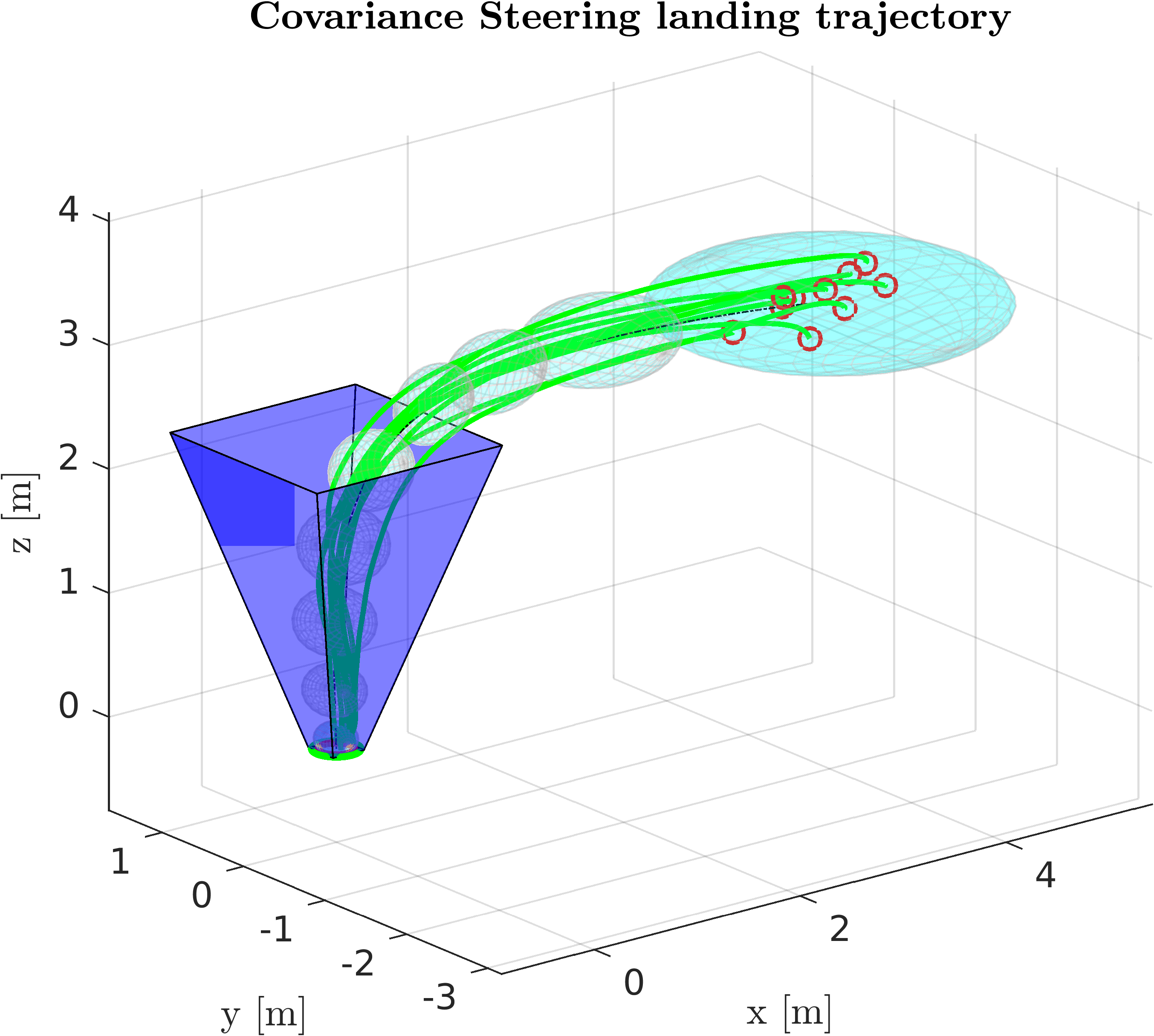}
    \caption{Landing trajectories with covariance steering controller + EKF aerodynamic compensator.}
    \label{fig: landing_OCS}
\end{figure}

For landing in the presence of winds, we solved a covariance steering problem where the state is constrained to lie in a feasible cone above the landing zone, described by equation \eqref{convex_constrained_cov} where the exact parameters can be found in Table \ref{cone_params}. 
We also include an input chance constraint of the form \eqref{convex_constrained_effort} with a gradual shrinking of the overall control authority as the vehicle approaches the ground.
This aims at preventing the quadrotor from doing aggressive corrections when close to the ground since these translate to large bank angles and less smooth landings.
The resulting trajectories are illustrated in Figure \ref{fig: landing_OCS}, while a comparison with state-of-the-art methods for the same problem is presented in Table \ref{tab:landing_comparison}.

\begin{table}[ht!] 
  \centering
    \caption{Parameters for the 4 chance constraint representing the sides of the cone illustrated in Figure \ref{fig:OCS_land} }
    
    \begin{tabular}{|c|c|c|c|c|c|c|}
    \hline
    \# & $\ell \t$                                         & $\alpha_x \t$                                             & $\beta_x$  \rule{0pt}{2.4ex} \rule[-1.2ex]{0pt}{0pt} \\ \hline
    1  & $ \left[ \quad 10.46,   \quad 0,  \quad 3.10    \quad 0_{3 \times 1} \right]  $   & $\left[ \quad 0, \quad  3.33, \quad 1, \quad 0_{3 \times 1} \right]$  & 0.5  \\ \hline
    2  & $ \left[ -10.46,        \quad 0,  \quad 3.10    \quad 0_{3 \times 1} \right]  $   & $\left[ \quad 0,       -3.33, \quad 1, \quad 0_{3 \times 1} \right]$  & 0.5  \\ \hline
    3  & $ \left[ \quad 0,       \quad 10.46, \quad 3.10  \quad 0_{3 \times 1} \right]  $   & $\left[ \quad 3.33, \quad  0, \quad 1, \quad 0_{3 \times 1} \right]$  & 0.5  \\ \hline
    4  & $ \left[ \quad 0,       -10.46,     \quad 3.10   \quad 0_{3 \times 1} \right]  $   & $\left[      -3.33, \quad  0, \quad 1, \quad 0_{3 \times 1} \right]$  & 0.5  \\
    \hline
  \end{tabular}
  \label{cone_params}
\end{table}

\begin{table}[ht!] 
    \centering
    \caption{Comparison with state-of-the-art for the Landing in the presence of wind benchmark.}
    \begin{tabular}{|c|c|c|c|}
    \hline
        & RMS $\;  \| \b x - \b x_{ref} \|_2 $ [cm] & rms $ \| \dot{\b \omega} \|_2 $ rad/s$^2$ \rule{0pt}{2.4ex} \rule[-1.2ex]{0pt}{0pt} \\ \hline
         OCS + EKF           & 6.0   & 10.1              \\ \hline
         OCS + INDI          & 5.6     & 12.0              \\ \hline
         LQR + EKF           & 4.6     & 16.0               \\ \hline 
         LRQ + INDI          & 3.6     & 17.1               \\ 
    \hline 
    \end{tabular}
    \label{tab:landing_comparison}
\end{table}

\section{Conclusion}
This paper presents a quadrotor control algorithm that optimally trades off some performance for flight smoothness, less control effort and actuator strain, while ensuring that the system performance constraints are met.
This is done by employing state-of-the-art results from Covariance Steering theory, a method used to compute optimal control policies for linear systems subject to Gaussian noise.  
Standard feedback linearization techniques are used to transform the quadrotor dynamics to those of a double integrator in combination with a novel aerodynamic force estimator that combines aerodynamic principles and machine learning, allowing it to be trained without data from wind tunnel testing, but rather from indoor flights at zero wind, where the aerodynamic disturbances are only due to the quadrotor motion.
A Kalman Filter is used to adapt the model with respect to external wind. 

\section*{ACKNOWLEDGMENT}
Support for this work has been provided by 
ONR award N00014-18-1-2828 and NASA ULI award \#80NSSC20M0163.
This article solely reflects the opinions and conclusions of its authors and not of any NASA entity. 
George Rapakoulias acknowledges financial support by the A. Onassis Foundation Scholarship.

\bibliographystyle{ieeetr}
\bibliography{george_refs}

\end{document}